\documentclass[preprint,10pt]{elsarticle}

\usepackage{amssymb}
\usepackage{amsthm}

\usepackage{amsfonts}
\usepackage{amsmath}

\usepackage[colorlinks]{hyperref}

\usepackage{slashbox}

\usepackage{adjustbox}
\usepackage[symbol]{footmisc}

\journal{arXiv}

\newtheorem*{lemma*}{Lemma}

\begin{document}

\begin{frontmatter}



\title{ Specular reflections removal in colposcopic images based on neural networks: Supervised training with no ground truth previous knowledge}


\author[add1]{Lauren Jimenez-Martin\footnote{Corresponding author at: Facultad de Matem\'atica y Computaci\'on. Universidad de la Habana. La Habana, Cuba.\\
	 E-mail address: l.jimenez@matcom.uh.cu (L. Jimenez-Martin)} }
\author[add1]{Daniel A. Vald\'es P\'erez}
\author[add2]{Ana M. Solares Asteasuainzarra}
\author[add1]{Ludwig Leonard }
\author[add1]{Marta L. Baguer D\'iaz-Romañach}

\address[add1]{Facultad de Matem\'atica y Computaci\'on. Universidad de la Habana. La Habana, Cuba}
\address[add2]{ Hospital Ginecobst\'etrico Universitario 'Ram\'on Gonz\'alez Coro'. La Habana, Cuba}

\begin{abstract}
Cervical cancer is a malignant tumor that seriously threatens women's health, and is one of the most common that affects women worldwide. For its early detection, colposcopic images of the cervix are used for searching for possible injuries or abnormalities. An inherent characteristic of these images is the presence of specular reflections (brightness) that make it difficult to observe some regions, which might imply misdiagnosis. In this paper, a new strategy based on neural networks is introduced for eliminating specular reflections and estimating the unobserved anatomical cervix portion under the bright zones. For overcoming the fact that the ground truth corresponding to the specular reflection regions is always unknown, the new strategy proposes the supervised training of a neural network to learn how to restore any hidden regions of colposcopic images. Once the specular reflections are identified, they are removed from the image, and the previously trained network is used to fulfill these deleted areas. The quality of the processed images was evaluated quantitatively and qualitatively. In 21 of the 22 evaluated images, the detected specular reflections were eliminated, whereas, in the remaining one, these reflections were almost completely eliminated. The distribution of the colors and the content of the restored images are similar to those of the originals. The evaluation carried out by a specialist in Cervix Pathology concluded that, after eliminating the specular reflections, the anatomical and physiological elements of the cervix are observable in the restored images, which facilitates the medical diagnosis of cervical pathologies. Our method has the potential to improve the early detection of cervical cancer.
\end{abstract}


\end{frontmatter}

\section{Introduction}
Cancer is a serious health problem due to its high incidence and mortality rates in the world. In particular, cervical cancer is one of the most common that affects women and is the fourth leading cause of female mortality from cancer worldwide \cite{factografico}. To increase the probability of successful treatment, early detection of the disease is necessary  \cite{robles1996introduction}. Before the appearance of cervical cancer, abnormal growth of squamous cells occurs in the cervical epithelium called cervical intraepithelial neoplasia. These precancerous and cancerous cells can be detected through a colposcopy, a visual inspection of the cervix by a clinical examination \cite{nazeer2011objective}. This test is performed by using a colposcope that captures color images outside the cervix. Once regions suggestive of intraepithelial lesion or cervical cancer have been identified, a targeted biopsy is performed to confirm the diagnosis, which contributes to developing treatment strategies in correspondence with the size and location of the lesions.

The cervix is a humid area and, when the light of the colposcope falls on it, specular reflections (SRs) may appear in the image.  Specular reflections raise challenging problems in medical image analysis, as it degrades (partially or entirely) the information in the affected pixels \cite{lange2005automatic}, which can lead to misdiagnosis. Therefore, it is imperative to find effective methods for eliminating the SRs and estimating the missing anatomical region under the bright zones. On this need, the present work is focused.

\subsection{Related work}\label{section:EstadodelArte}
Previous studies on colposcopic image processing have allowed the detection of SR regions \cite{das2011elimination, meslouhi2011automatic, Palmer}. Once these regions are identified,  SRs removal can be treated as an inpainting problem, which consists of filling in the missing regions based on the remaining image data \cite{siavelis2020improved}.
The restored region must be consistent with the cervix anatomy. Different approaches have been taken to deal with SRs removal by inpainting methods in colposcopic images.
The authors of \cite{lange2005automatic} proposed filling in the affected regions by interpolating the RGB (Red, Green, Blue) color components individually from the surrounding regions based on Laplace's equation and modifying the intensity component of the HSI (Hue, Saturation, Intensity) color space transformed image. In  \cite{zimmerman2006automatic}, it was assumed that the highlights formed on the moist surface of the cervix are very small and the color underneath each highlight is nearly constant and similar to the color of the pixels in the immediate surroundings. So, it was proposed to fill in the SR regions by propagating the surrounding color information. 
After the color value of the detected SR pixels is set to zero, an iterative process replaces each pixel value inside the SR region by the mean color of its non-zero neighbors. A similar idea was followed in \cite{xue2007comparative}, but the pixels inside the SR region were replaced by the weighted color values of their neighboring pixels based on the average gradient direction of the SR region.
All these methods are based on the gradual propagation of colors from edges toward the specular reflection center, and they provide satisfactory results when applying to small areas. 
In contrast, \cite{ meslouhi2011automatic} argues that SR regions are typically large, so reconstructing missing information by only considering neighboring pixel value is not realistic.
They suggest applying the multi-resolution inpainting technique proposed in \cite{shih2005digital} to restore the SRs by blocks with different brightness levels and applying histogram equalization to homogenize each restored block and reduce the effect of dividing the SRs into several blocks.
By considering the colposcopic image with SRs as a matrix with unknown entries, \cite{Danilo} proposes estimating SR regions employing Non-Negative Matrix Factorization. However, the quality of the reconstruction strongly depends on the initial parameters of the algorithm. The authors of \cite{wang2019detection} proposed inpainting of SRs in colposcopic images with an exemplar-based method.

In recent years, there has been an increasing amount of literature on Convolutional Neural Networks (CNNs) to perform image inpainting tasks. CNNs are used as a feature extraction method through the process of convolution.
Particularly in medical image restoration, the use of neural networks to solve inpainting problems has been increasing due to the good performance they have shown on images from other domains. 
The authors of \cite{ ali2019deep} chose to incorporate YOLOv3 with spatial pyramid pooling (YOLOv3-spp) for robust detection and improved inference time for endoscopic artifacts, which may affect the physician's visual assessment, and propose the use of a Conditional Generative Adversarial Network (CGAN) to restore the affected areas in the image. 
The use of CNN combined with adversarial training \cite{goodfellow2014generative} has produced excellent results on inpainting tasks, with perceptual similarity to the original image.
The authors of \cite{murugesan2019recon} propose a new architecture based on generative adversarial networks, Reconstruction Global-Local GAN (Recon-GLGAN), for magnetic resonance image reconstruction. However, despite the good performance of the deep learning algorithms in these inpainting tasks, they have not been applied to specular reflection removal in colposcopic images.

In this paper, we present a new approach to eliminate specular reflections based on training a network to learn how to restore any hidden region of colposcopic images. Once the SRs are identified, they are removed from the image, and the previously trained network is used to fulfill these deleted areas. We use a convolutional denoising autoencoder trained for the completion task using a  small database, contrary to the belief that, for optimal performance, large training datasets are needed for models based on deep architectures.

This article is organized as follows. The next section describes the neural network architecture selected to eliminate specular reflections in colposcopic images, and details the strategy followed for its training. 
The fourth section presents the experimental process carried out as well as a qualitative and quantitative analysis of the obtained results.

\section{Material and methods} \label{section:Diseño Experimental}
\begin{figure}[htb]%
	\begin{center}
		\includegraphics[scale=0.4]{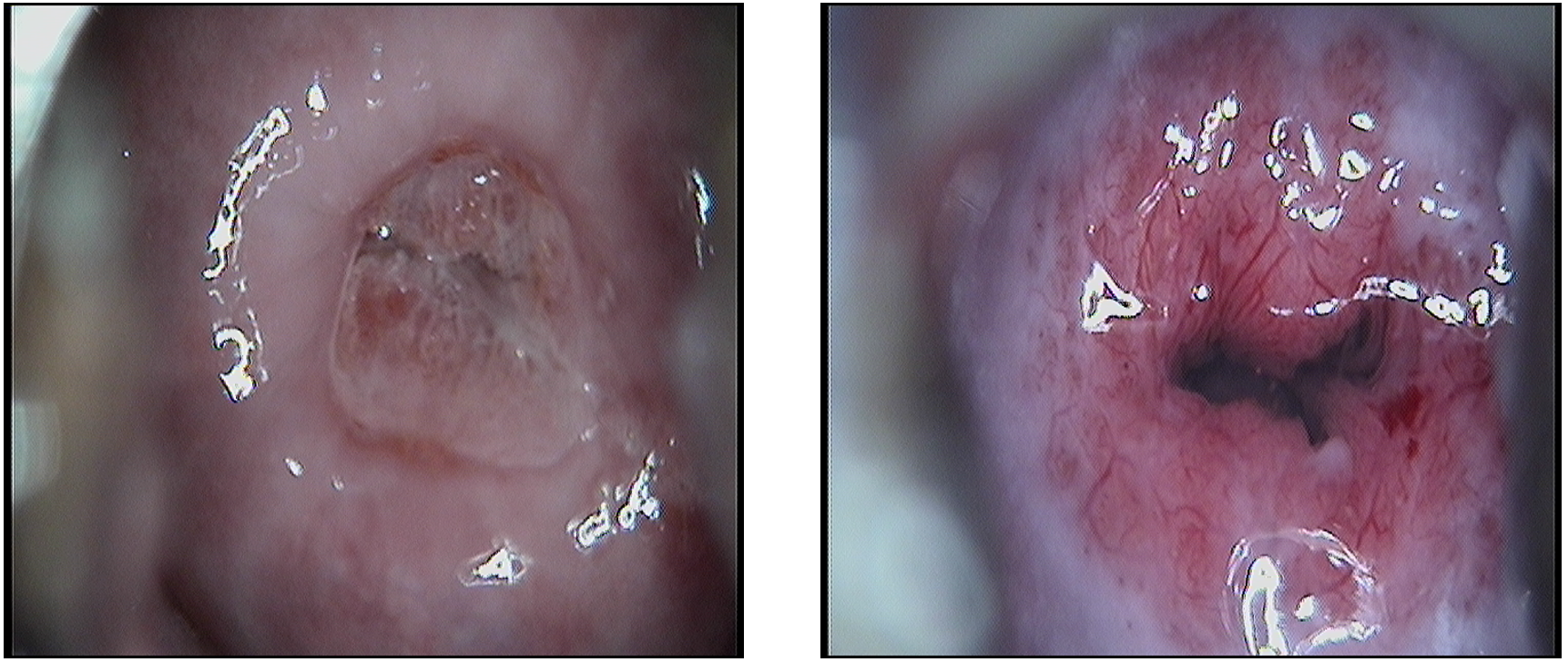}
	\end{center}
	\caption{Colposcopic images with SRs from different patients. \label{fig:comparacion_brillo_img_diferentes}}%
\end{figure}

This section deals with the specular reflections removal in colposcopic images using deep learning techniques. 
Figure \ref{fig:comparacion_brillo_img_diferentes} presents two typical colposcopic images from different patients showing large differences in color and shape of the cervices and the presence of scattered brightness (SRs).
After considering the main limitation of the standard supervised learning for eliminating SRs in a colposcopic image, a new strategy for applying a supervised learning algorithm is introduced despite not knowing the ground truth of the problem to be solved.

\subsection{Characterization and detection of specular reflections}\label{sec:Palmer}

Pixels belonging to regions with SRs are characterized by high intensity (\textit{Int}) and low color saturation (\textit{Sat}) \cite{T.M.Lehmann-C.Palm}.
These characteristics allow a preliminary identification of such regions by applying a threshold criterion to these image values \cite{zimmerman2006automatic}.
The authors of \cite{Palmer} studied different approaches to detect SR regions on Cuban colposcopic images. They proposed an algorithm based on the application of thresholds to the maximum intensity $ Int_{max} $ of the image regardless of color saturation. The pixels with intensity higher than the thresholds were classified as SR. This algorithm was chosen to detect the SR region in this work.

\subsection{Peculiarities of the problem to be solved and feasibility of supervised training}\label{sec:training_clasico}\vspace{0.25cm}

\textbf{Formulation of the problem to be solved}. \textit{Let us denoted by
	\begin{itemize}
		\item $ A  $ the area of the cervix focused with the colposcope
		\item $ I $ the image of A captured by the colposcope, i.e., image with SR regions.
		\item $ I_e $ the ideal image showing the complete content of A, i.e., image without the SR regions.
	\end{itemize}
	Target: From image $ I $, obtain image $ I_e $ using supervised learning.}\\

The supervised training of a neural network depends on knowing in advance the training set consisting of $ N $ pairs ($ x_{i} $, $ y_{i} $) of inputs and outputs, $ i=1,...,N $. Adjusting this to the problem to be solved, the training set would have as input $ x_i $ a colposcopic image $ I $ and as output (ground truth) $ y_{i}$  the image $ I_e $. 

\begin{figure}[htb]%
	\begin{center}
		\includegraphics[scale=0.4]{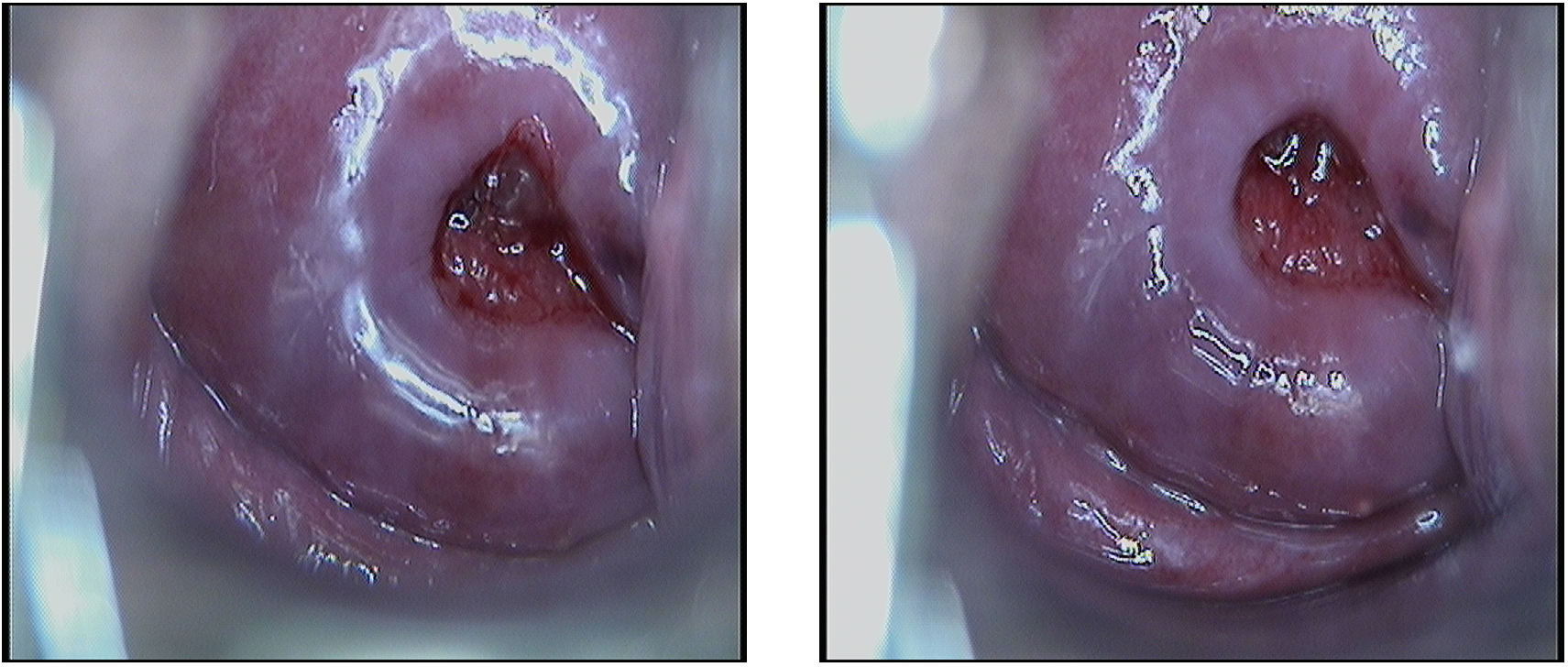}
	\end{center}
	\caption{Colposcopic images with SRs of the same patient. \label{fig:comparacion_brillo_misma_paciente}}%
\end{figure}

However, the presence of SRs is an inherent characteristic of colposcopic images produced by the reflection of the colposcope light on the wet areas of the cervix. Since the areas and the  humidity level are different for each patient, the distribution of SRs in the images is heterogeneous (see Figure \ref{fig:comparacion_brillo_img_diferentes}).  Moreover, for a patient, the incidence of the colposcope light on an area $ A $ at different angles may result in images I with SRs located on different regions, but it would never result in an image Ie showing the complete content of A, as seen in Figure \ref{fig:comparacion_brillo_misma_paciente}. Therefore, for image I of the area A, the corresponding ground truth $ I_e $ is unknown. For this reason, applying supervised training of a neural network directly to solve this problem is not feasible.

\subsection{Strategy for applying a supervised learning algorithm}  \label{sec:entrenamientoPropuesto}
Considering the above-mentioned peculiarities of the problem to solve, in this section, the problem of SRs removal in colposcopic images using supervised learning is reformulated, the architecture and training of the network defined, and the use of the network to solve the problem described.\vspace{5.0cm}

\textbf{Reformulation of the problem to solve}. \textit{Let us denoted by
	\begin{itemize}
		\item $ A  $ the area of the cervix focused with the colposcope
		\item $ I $ the image of A captured by the colposcope, i.e., image with SR regions.
		\item $ I^{'}  $ and $ I^{''} $  modifications of image $ I $ such that there are hidden regions in the image $ I^{''} $ that are known in the image $ I^{'} $.
\end{itemize}}
\textit{Targets:
	\begin{enumerate}
		\item Train a network for learning to complete the hidden content in $ I^{''} $ by trying to obtain $ I^{'} $. 
		\item Use the trained network to complete the regions with SRs in a colposcopic image $ D $, which solves the original problem.
\end{enumerate}}

Since the network is trained to restore any hidden region of colposcopic images, it is expected that the network will be able to reconstruct any unobserved anatomical cervix portion under the SR regions.   
\newline

\textbf{Image pre-processing. }
The modified images $ I^{'}  $ and $ I^{''} $ for constructing the data set of the above-reformulated problem are constructed as follows: 

\begin{enumerate}
	
	\item The regions with SRs are identified using the algorithm mentioned in Section \ref{sec:Palmer}, and a binary $m \times n$ mask $ M_r $ (real mask) was associated with them. 
	The entry $(i,j)$ of the real mask of a colposcopic image $ I_{m \times n} $ is defined as	
	
	\begin{equation*}
	[M_r]_{ij}= \left\{ \begin{array}{lcc}
	0 &  \mbox{if the  pixel }I_{ij}  \mbox{ has  SR  }\\
	1 &  \text{otherwise }\\
	\end{array},
	\right.
	\end{equation*}

	where $ i = 1,...,m $;
	$ j = 1,...,n $. The image $ I^{'} = I * M_r $ is constructed, where the symbol $ * $  denotes the Hadamard product for two matrices. See Figure \ref{fig:ejemplo_entrenamiento_propuesto} center.
	
	\item From the regions without SRs of $ I $, regions of interest  will be selected as hidden regions (HR), and a new $m \times n$ mask $ M_h $ (hidden mask) will be associated with them. That is, 
	\begin{equation*}
	[M_h]_{ij}= \left\{ \begin{array}{lcc}
	
	0 &  \mbox{if the  pixel } I^{}_{ij} \in  HR \\
	1 &  \text{otherwise }\\
	\end{array},
	\right.
	\end{equation*} 
	where $ i = 1,...,m $;	$ j = 1,...,n $. 
	Then, $ I^{''} = I^{'} * M_h  $ is constructed.
	See Figure \ref{fig:ejemplo_entrenamiento_propuesto} right.

	\begin{figure}[htb]%
		\begin{center}
			\includegraphics[scale=0.55]{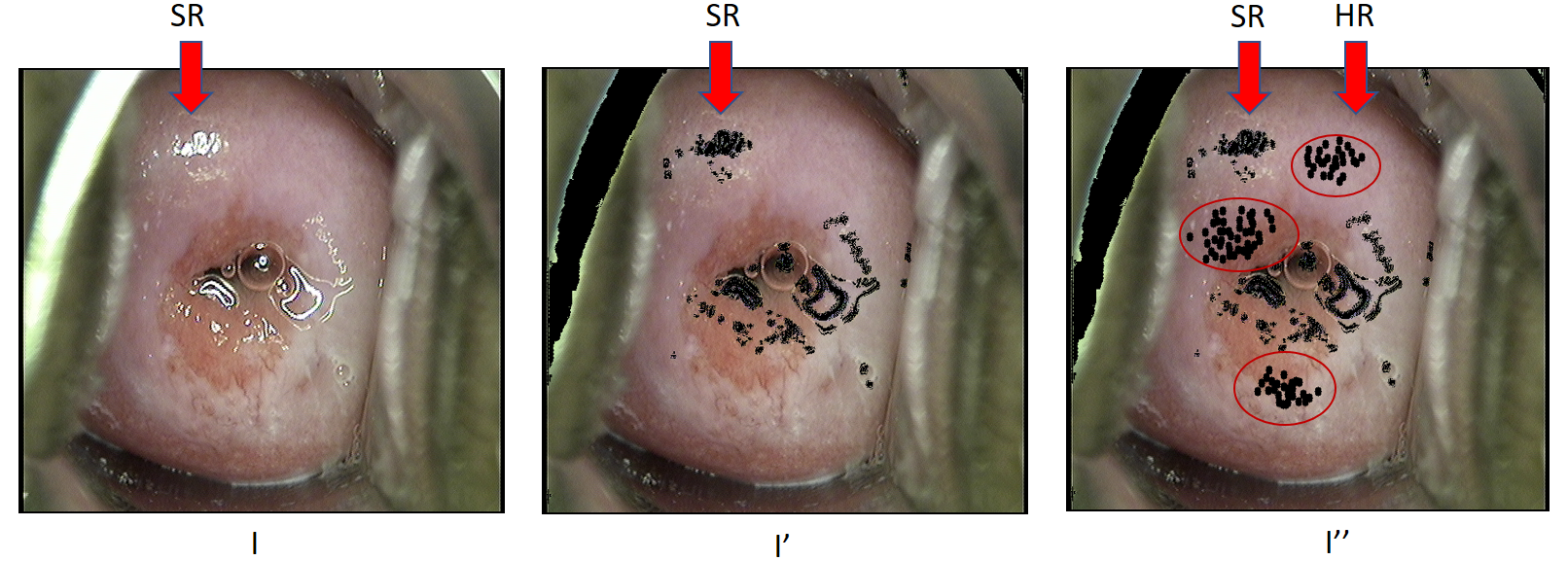}
		\end{center}
		\caption{Example of the modified images $ I^{'} $ and $ I^{''}$ of $I$. The black regions marked inside the red circles in $ I^{''} $ are the hidden regions ($ HR $).
			\label{fig:ejemplo_entrenamiento_propuesto}}%
	\end{figure}
	
	\item Applying steps 1 and 2 on a set of $ N $ colposcopic images $ I_{i} $, we conform the tuple $ ({I_i}^{''}, {I_i}^{'}) $ for the data set, where $ i = 1,...,N $.
	
\end{enumerate}

\subsubsection{ Architecture of the neural network}\label{sec:arquitectura}
The authors of  \cite{iizuka2017globally} propose a Generative Adversarial Network (GAN) model that improves the results obtained by  \cite{barnes2009patchmatch,huang2014image,pathak2016context} for solving image inpainting problems arisen in different (non-medical) domains such as faces and landscapes. The model architecture by \cite{iizuka2017globally} comprises three networks: a generator, a global context discriminator, and a local context discriminator. The generative network is fully convolutional and is used to fill in the missing regions of the image, while the global and local context discriminators are auxiliary networks used exclusively for training.
Unlike other approaches focused on image generation, their method does not generate images from noise.

To solve the reformulated problem specified in Section \ref{sec:entrenamientoPropuesto}, we will use a network architecture based on the generative network architecture of \cite{iizuka2017globally}.   
This network follows an encoder-decoder structure that initially decreases the image resolution before its further processing, reducing memory usage and computation time. Subsequently, the network output is restored to the original resolution using deconvolution layers (the opposite process of a convolution filter). The resolution is decreased twice using convolutions of stride 1/4 of the original size, which is important to generate a non-blurred texture in the missing regions \cite{iizuka2017globally}. Using
dilated convolutions at lower resolutions, the model can effectively process larger input image areas when computing each output
pixel than with standard convolutions \cite{iizuka2017globally}. The model also has batch normalization layers after all convolutional layers except for the last one.

\begin{table}[h!tb]
	\centering
	\begin{tabular}{lllll}
		\hline\hline
		Type & Kernel & Dilation & Stride & Filters\\ \hline
		convolution & \multicolumn{1}{c}{5$\times $5} & \multicolumn{1}{c}{1} & 
		\multicolumn{1}{c}{1$\times $1} & \multicolumn{1}{c}{32} \\ \hline
		convolution & \multicolumn{1}{c}{3$\times $3} & \multicolumn{1}{c}{1} & 
		\multicolumn{1}{c}{2$\times $2} & \multicolumn{1}{c}{64} \\ 
		convolution & \multicolumn{1}{c}{3$\times $3} & \multicolumn{1}{c}{1} & 
		\multicolumn{1}{c}{1$\times $1} & \multicolumn{1}{c}{64} \\ \hline
		convolution & \multicolumn{1}{c}{3$\times $3} & \multicolumn{1}{c}{1} & 
		\multicolumn{1}{c}{2$\times $2} & \multicolumn{1}{c}{128} \\ 
		convolution & \multicolumn{1}{c}{3$\times $3} & \multicolumn{1}{c}{1} & 
		\multicolumn{1}{c}{1$\times $1} & \multicolumn{1}{c}{128} \\ 
		convolution & \multicolumn{1}{c}{3$\times $3} & \multicolumn{1}{c}{1} & 
		\multicolumn{1}{c}{1$\times $1} & \multicolumn{1}{c}{128} \\ 
		dilated convolution  & \multicolumn{1}{c}{3$\times $3} & 
		\multicolumn{1}{c}{2} & \multicolumn{1}{c}{1$\times $1} & \multicolumn{1}{c}{
			128} \\ 
		dilated convolution  & \multicolumn{1}{c}{3$\times $3} & 
		\multicolumn{1}{c}{4} & \multicolumn{1}{c}{1$\times $1} & \multicolumn{1}{c}{
			128} \\ 
		dilated convolution & \multicolumn{1}{c}{3$\times $3} & 
		\multicolumn{1}{c}{8} & \multicolumn{1}{c}{1$\times $1} & \multicolumn{1}{c}{
			128} \\ 
		dilated convolution & \multicolumn{1}{c}{3$\times $3} & 
		\multicolumn{1}{c}{16} & \multicolumn{1}{c}{1$\times $1} & 
		\multicolumn{1}{c}{128} \\ 
		convolution & \multicolumn{1}{c}{3$\times $3} & \multicolumn{1}{c}{1} & 
		\multicolumn{1}{c}{1$\times $1} & \multicolumn{1}{c}{128} \\ 
		convolution & \multicolumn{1}{c}{3$\times $3} & \multicolumn{1}{c}{1} & 
		\multicolumn{1}{c}{1$\times $1} & \multicolumn{1}{c}{128} \\ \hline
		deconvolution & \multicolumn{1}{c}{4$\times $4} & \multicolumn{1}{c}{1}
		& \multicolumn{1}{c}{1/2$\times $1/2} & \multicolumn{1}{c}{64} \\ 
		convolution & \multicolumn{1}{c}{3$\times $3} & \multicolumn{1}{c}{1} & 
		\multicolumn{1}{c}{1$\times $1} & \multicolumn{1}{c}{64} \\ \hline
		deconvolution & \multicolumn{1}{c}{4$\times $4} & \multicolumn{1}{c}{1}
		& \multicolumn{1}{c}{1/2$\times $1/2} & \multicolumn{1}{c}{32} \\ 
		convolution & \multicolumn{1}{c}{3$\times $3} & \multicolumn{1}{c}{1} & 
		\multicolumn{1}{c}{1$\times $1} & \multicolumn{1}{c}{16} \\ 
		output & \multicolumn{1}{c}{3$\times $3} & \multicolumn{1}{c}{1} & 
		\multicolumn{1}{c}{1$\times $1} & \multicolumn{1}{c}{3} \\ \hline\hline
	\end{tabular}
		\caption{ Architecture of the network. The activation function of each layer is a Rectified Linear Unit (ReLU), except the last one. The last layer (output) is a convolutional layer with a sigmoid function to normalize the output to the [0, 1] range.}
	\label{tab:arquitectura}
\end{table}

\begin{figure}[htb]%
	\begin{center}
		\includegraphics[scale=0.55]{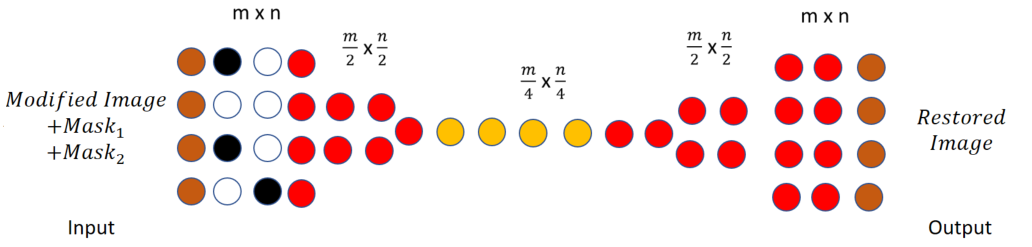}
	\end{center}
	\caption{Diagram of the network architecture.\label{fig:modelo}}%
\end{figure}

Specifically, we use the above-described network architecture with the following modifications.
The number of filters used in each layer is reduced by half to decrease the number of operations to be performed and the computation time. Table \ref{tab:arquitectura} shows the final architecture of the network. Instead of an RGB image with a mask representing the region to be filled, the model input is an RGB image with two masks. The first mask identifies the black pixels of the input image that should be retained in the output image. The second mask identifies the black pixels of the input image that must be restored. Both masks are binary arrays of the image size, where the positions with value 0 will be those representing the black pixels, and 1 the rest of the image.
It is intended that, from these patterns, the network learns which areas to copy and which ones to restore. A general representation of this model can be seen in Figure \ref{fig:modelo}.

To restore hidden regions of a colposcopic image $ I $  the input image is $ I^{''}$ defined in Section \ref{sec:entrenamientoPropuesto}. The first mask is the real $ M_{r} $ of $ I $, and the second is the hidden $ M_{h} $.
On the other hand, to restore anatomical cervix portion under the SR regions in a colposcopic image $ I $, the input image is $ I^{'} $, the first mask is composed of a matrix full of 1, and the second mask is the real $ M_{r} $ of $ I $ representing the pixels to be restored. 
\vspace{3.0cm}

The optimizer used in training is the  Adadelta algorithm \cite{Zeiler}, which automatically sets a learning rate for each weight in the network. Adadelta optimization is a stochastic gradient descent method that is based on adaptive learning rate per dimension to address two drawbacks:
\begin{enumerate}
	\item The continual decay of learning rates throughout training, implying an incorrect update of the weights. In the worst case, this may prevent the neural network from continuing its training \cite{pascanu2013difficulty}.
	
	\item The need for a manually selected global learning rate \cite{Zeiler}.
\end{enumerate}
This algorithm adapts learning rates based on a moving window of gradient updates, rather than accumulating all previous gradients. In this way, Adadelta continues to learn even when many updates have been made.

The loss function used for training the network is the Mean Squared Error (MSE)
\begin{center}
	$ || h_{\theta}(x) - y||_2$
\end{center}
defined by the Euclidean norm $ || \cdot||_2 $, the obtained output $ h_{\theta}(x) $ of the network for the input $ x $, and the expected output $ y $. Since in our case we intend to compute the distance between RGB images, we use the objective function 
\begin{equation}\label{eq:mse}
\frac{1}{m\times n\times 3}  \sum_{i=1}^{m} \sum_{j=1}^{n} \sum_{k=1}^{3} (h_{\theta}( I^{''}_{ijk}) - I^{'}_{ijk})^2.
\end{equation}

\subsubsection{Image selection and hidden regions }\label{Selección_de_parámetros}

We use a colposcopic imaging database from the  University  Gynecobstetric  Hospital 'Diez de Octubre'  and the University  Gynecobstetric Hospital 'Ram\'on Gonz\'alez Coro'. This database contains several images (from different angles) of each patient. The images were partitioned into three sets, training, validation, and test so that they did not share images of the same patient. This partition guarantees the independence between the data of the different sets.
\vspace{3.0cm}

\begin{figure}[htb]%
	\begin{center}
		\includegraphics[scale=0.45]{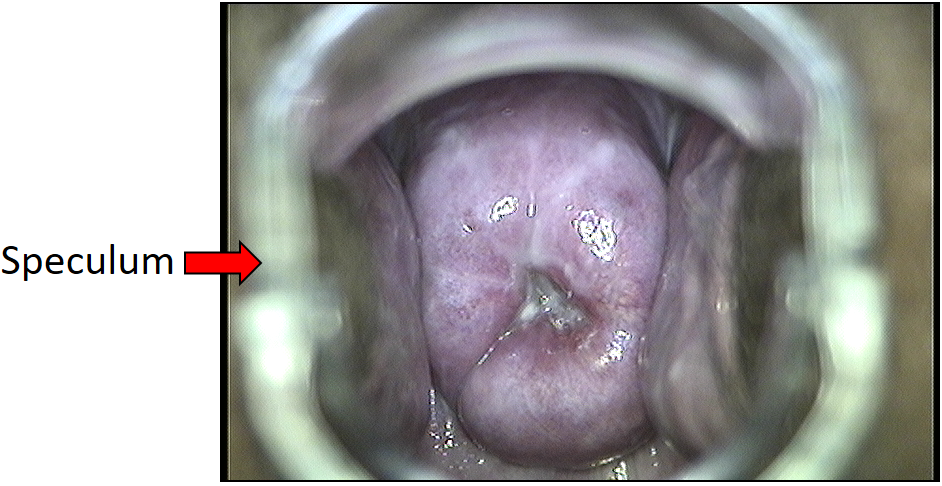}
	\end{center}
	\caption{Example of a colposcopic image containing part of the speculum. \label{fig:especulo}}%
\end{figure}

Frequently, colposcopic images may contain part of the speculum\footnote{Medical instrument that holds open the entrance orifices of different body cavities such as the vagina to perform examinations.} (see Figure \ref{fig:especulo}) or areas outside the cervix that are not subject to clinical studies. For delimiting the cervical area in this type of image, several segmentation methods have been proposed \cite{lange2005automatic2,li2007automated,bai2018automatic}. However, they assume that the  SR regions have been previously eliminated, so it is not realistic to use them in this research. Since hidden regions must be inside the gynecological interest areas, they were selected manually by visual inspection. 
The manual selection was carried out with the help of a program that displays the images and allows selecting the hidden pixels. The program also automatically generates the hidden mask $ M_h $ corresponding to the previously selected HRs.\newline

As mentioned above, the bright pixels do not have a fixed distribution and concentration in the images (see Figures \ref{fig:comparacion_brillo_img_diferentes} and \ref{fig:comparacion_brillo_misma_paciente}). Therefore, the shape and concentration of the hidden regions were created heterogeneously. To ensure a varied representation of different features of the cervix during the network training process, the pixels of the hidden regions were chosen from distinctive regions of the cervix displayed in the images (as those with lesions, blood vessels, distinct textures, and clean areas) as well as from other areas of the images randomly selected.

\subsection{Implementational and computational issues}
The strategy proposed in Section \ref{sec:entrenamientoPropuesto} was implemented in Python 3.7 with the use of the deep learning package Keras on top of the machine learning platform TensorFlow. In addition, Google Colaboratory (also known as Colab) was used to accelerate the training process. The types of GPUs available in Colab often include Nvidia K80s, T4s, P4s, and P100s, but they vary over time.
\vspace{1cm}

\section{Experimentation, Results and Discussion}\label{section:experiments}

When initial values of the weights of a neural network are taken randomly, two networks trained with the same data set and the same number of epochs may result in networks with different final weights, thus with different generalization errors. 
For this reason, 16 networks were trained with the architecture proposed in Section \ref{sec:arquitectura}, the same data set, and 240 epochs. With the purpose of restoring hidden regions, 120 images from the colposcopic image database were selected to construct the training set, 20 for the validation set and 22 for the test set.
The last 22 images were also used to create the real test set for evaluating the performance of the trained network to restore SR regions.
These sets were arranged in the way explained in Section \ref{sec:entrenamientoPropuesto}.


\subsection{Performance of the trained networks to restore hidden regions of colposcopic images}\label{sec:Problema_aumentado}
\begin{figure}[htb]%
	\begin{center}
		\includegraphics[scale=1.5]{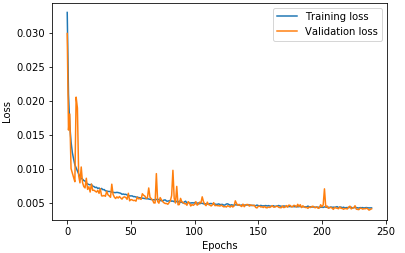}
	\end{center}
	\caption{\textbf{R3} network learning curves. \label{fig:learning_curves_of_R3}}%
\end{figure}

An identifier \textbf{R}\textit{x}  was assigned to each neural network, where $ x $ is the number associated with the network. Table \ref{tab:errores} shows, in increasing order, the validation errors corresponding to each network. Taking into account that the trained network \textbf{R3} presents the lowest validation error, it was selected to restore the hidden regions. Figure \ref{fig:learning_curves_of_R3} shows \textbf{R3} learning curves. It can be appreciated that there is no overfitting as training and validation curves behave similarly. To evaluate the performance of the selected network regarding the others, a series of qualitative and quantitative comparisons are carried out.

\begin{table}[h!tb]
	\centering
	\begin{adjustbox}{width=1\columnwidth,center}
		\begin{tabular}{ | l | l | l | l | l | l | l | l | l | l | l | l | l | l | l | l | l|  p{0.1cm} |}
			\hline		
			ID & R3	& R14 & R4 & R2	& R6 & R10	& R1	& R7 & R5 &	R12	& R15 &	R11	&		R8	& R0 & R13	& R9 \\ \hline
			VE $ \times10^{-3} $  & 4.12
			& 4.34 	& 4.41 & 4.52 
			& 4.59 & 4.60 & 4.66 & 4.68 & 4.76 & 4.96 & 5.16 & 5.37 & 5.44 & 5.46 & 5.50 & 5.77\\\hline
		\end{tabular}
	\end{adjustbox}
	\caption{ List of trained networks, identified by an ID, and their corresponding validation error (VE). \newline}
	\label{tab:errores}
\end{table}

\begin{figure}[h!tb]%
	\begin{center}
		\includegraphics[scale=0.95]{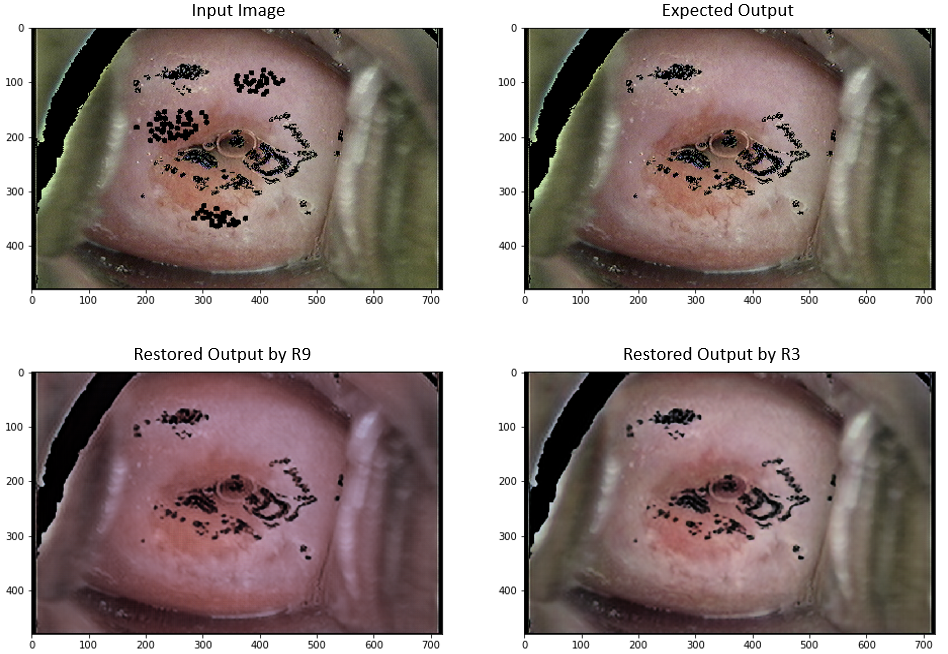}
	\end{center}
	\caption{Comparison of the restored images obtained by the networks \textbf{R9} and \textbf{R3} with the highest and lowest validation error, respectively. \label{fig:comapracion_R2_R3}}%
\end{figure}

By visual inspection of the colposcopic images restored by the different networks, certain qualitative differences can be observed, for example, in the color tonality. 
Figure \ref{fig:comapracion_R2_R3} shows a comparison of the restored images obtained by the networks with the lowest R3 and highest R9 validation errors
(see Table \ref{tab:errores}). \newline

\begin{figure}[h!tb]%
	\begin{center}
		\includegraphics[scale=0.51]{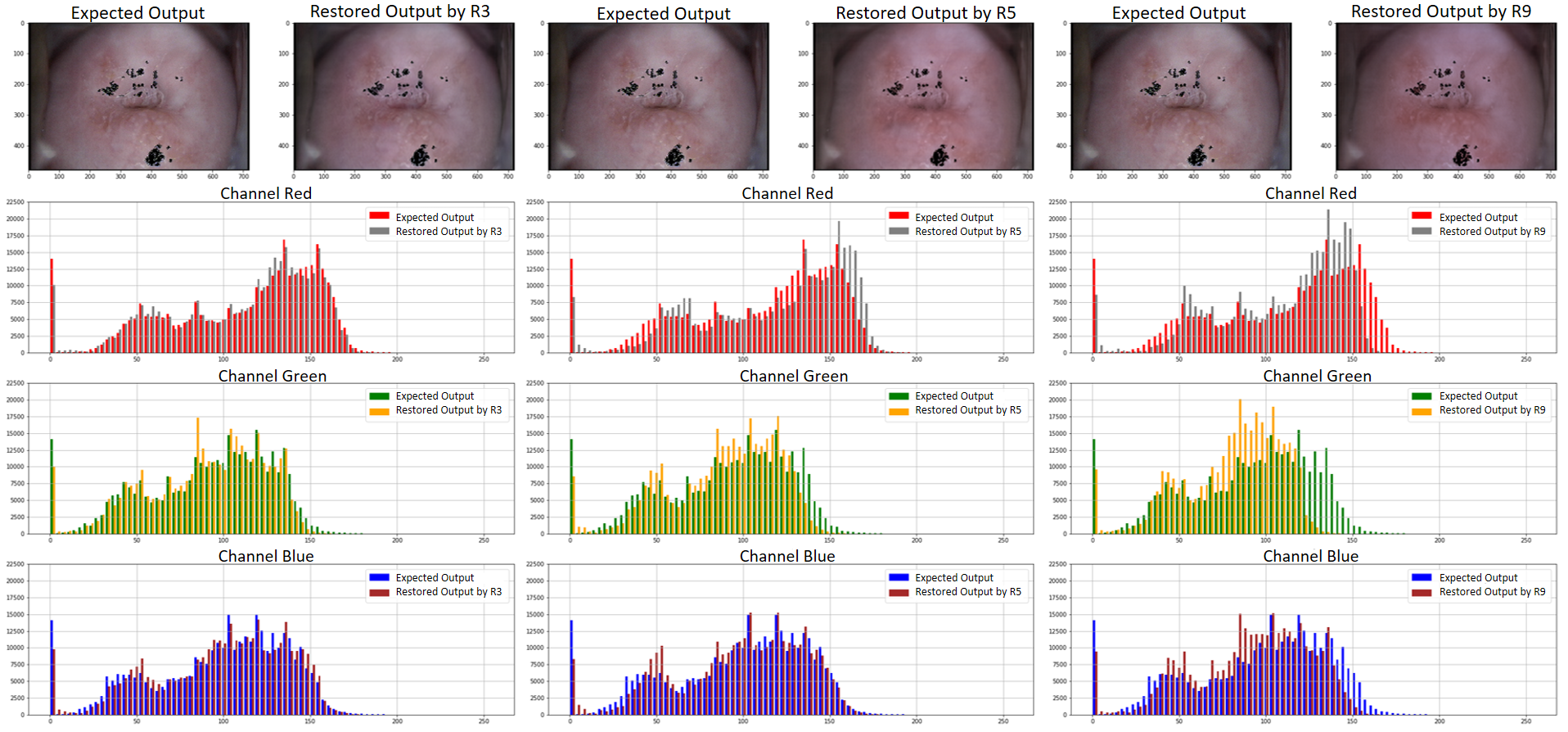}
	\end{center}
	\caption{Result obtained after completing a colposcopic image using the network with lowest (\textbf{R3}, left), medium (\textbf{R5}, center), and highest (\textbf{R9}, right) validation error. The top row shows the expected and restored output images for each network, while the remaining rows show the histograms of color intensities per channel corresponding to the expected output image (in red, green, and blue) and obtained by each network (gray, yellow, brown).	\label{fig:176-12_R3-R5-R9}}%
\end{figure}

The color of a pixel in a colposcopic image depends on the combination of the intensities of its three channels (RGB). To analyze the restoration performed on each channel, the histograms of their pixel intensities were compared independently. Figure \ref{fig:176-12_R3-R5-R9}, top row, shows the results obtained when restoring the same image with the networks \textbf{R3}, \textbf{R5}, and \textbf{R9}  (extreme and central values of Table \ref{tab:errores}). In the remaining rows of the Figure, histograms of the pixel intensities corresponding to expected output and obtained output images for each network, separated by channels, are superimposed. It can be seen that there is a considerable variation in the estimated pixel values of each channel among the three networks. The distribution of intensities estimated in each channel by the \textbf{R3} network is the closest to the expected distribution. By considering the whole test set of images, the test errors of the networks \textbf{R3}, \textbf{R5}, and \textbf{R9} are 0.0037 $ \pm $0.0007, 0.0045 $ \pm $0.0008, and 0.0058 $ \pm $0.0009, respectively, at the 95$\%$ of significance level. Observe that the confidence intervals [0.0030,0.0044] and [0.0049,0.0067] for the error of the networks \textbf{R3} and \textbf{R9} on the test set have null intercept, meaning that there is a significant difference in their capacity of generalization, i.e., of restoration of hidden regions of the colposcopic images.

\begin{figure}[h!tb]%
	\begin{center}
		\includegraphics[scale=0.71]{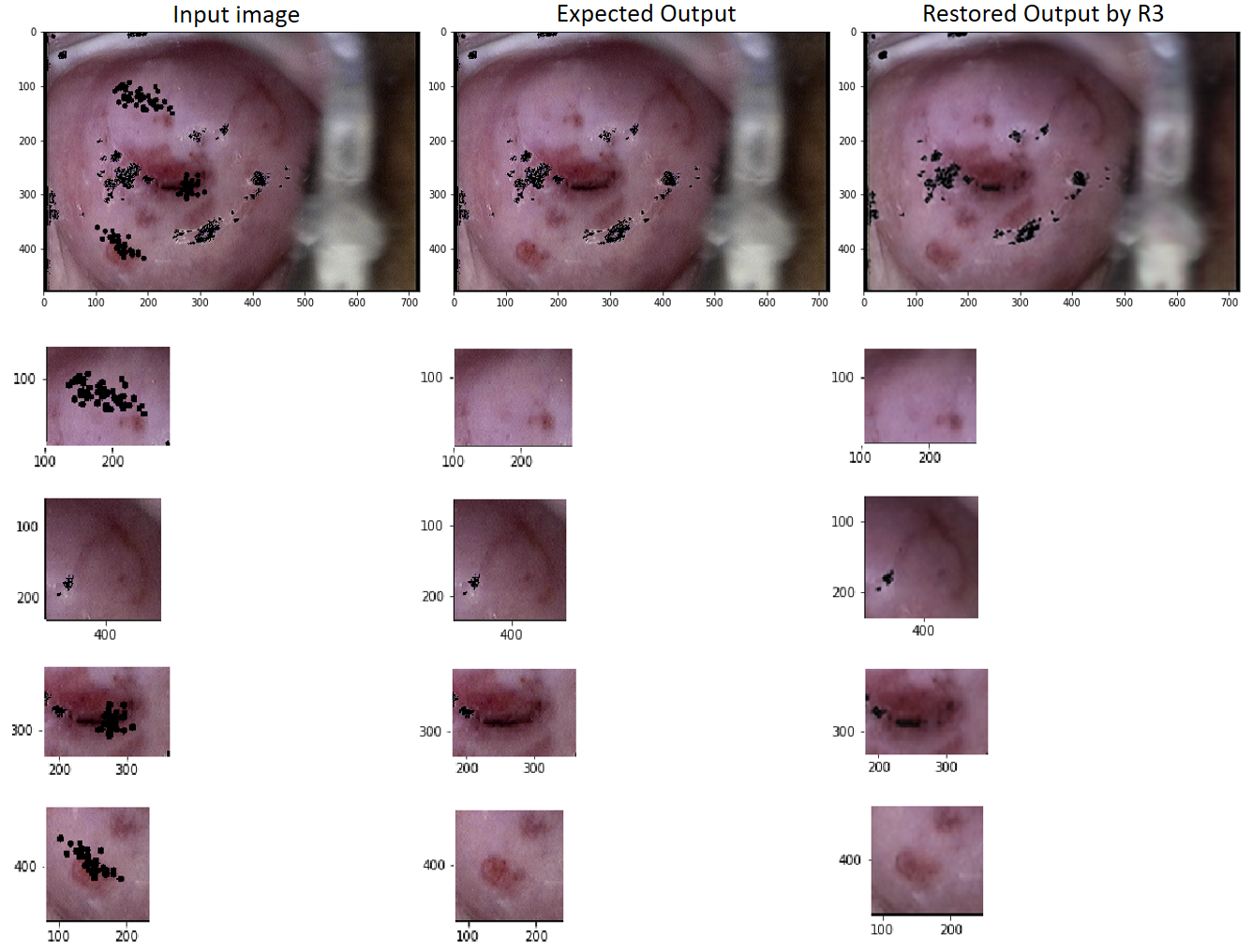}
	\end{center}
	\caption{Result obtained after restoring a colposcopic image using the trained network \textbf{R3} with the lowest validation error. From left to right, the top row shows the input, expected output, and obtained output images, respectively. The remaining rows show some areas of interest of the image on the top of the column. The axes of these areas indicate their position within the corresponding image.			   \label{fig:restauracion_aumentada_R3}}%
\end{figure}

In what follows, the performance of the trained network \textbf{R3} to restore the hidden regions in images of the test set is analyzed in detail.

Figure \ref{fig:restauracion_aumentada_R3} shows fine details in the estimation of various distinctive regions of a restored image. The top row presents the input image to the network (left), the expected output of the network (center), and the output obtained from the network (right). 
The remaining rows show areas of interest within the images as mentioned above, emphasizing some of their characteristics. These areas are shown in the column corresponding to their image.  Rows 2, 4, and 5 contain hidden regions and their respective restorations. Row 3 enhances how the tissue features outside the unknown regions are maintained during the restoration.

\begin{figure}[h!tb]%
	\begin{center}
		\includegraphics[scale=1]{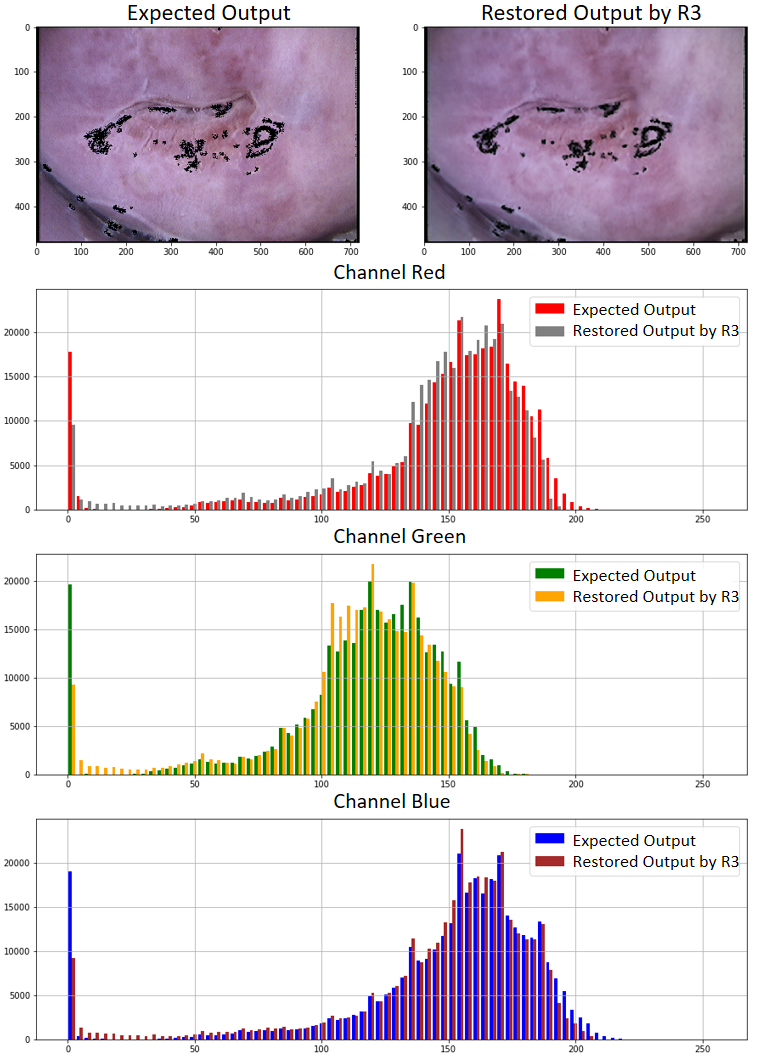}
	\end{center}
	\caption{Result obtained after restoring a colposcopic image using the network with the lowest validation error \textbf{R3}. The top row shows the expected output image and obtained output image. The remaining rows show the histograms of color intensities per channel corresponding to the expected (red, green and blue) and obtained  (gray, yellow, brown) output images. \label{fig:R3_hist_por_canales_167-2}}%
\end{figure}

Figure \ref{fig:R3_hist_por_canales_167-2}, top row, presents a restored image by the network \textbf{R3} and its corresponding expected image, whereas  their histograms of the pixel intensity per channel are shown in the remaining rows. Observe the suitable reproduction of the expected intensities by the network \textbf{R3} for all the intensity values higher than 0. Note also that there is a higher error when restoring the pixels with intensity 0 in the three channels of the images.
Looking in detail at the black areas in the expected output image (EO) and comparing them with the corresponding areas in the restored output image (RO), some scattered pixels can be seen around the black areas in EO, but not in RO. This reveals that black pixels (intensity 0 in the three channels) take color (intensities between 0 and 255), and vice versa, pixels with colors become black. This is a predicted result that reveals certain impressions of the network to restore the color of the pixels in the abrupt border between the black and the colored regions. However, as seen from the images on the top row of the Figure, these inaccuracies do not produce some appreciable visual distortion in the restored image.

As mentioned in Section \ref{sec:arquitectura}, the trained networks use as loss function the mean square error (\ref{eq:mse}), which quantifies the average of the errors between the three channels of the obtained image and the expected output image. However, this does not imply that the performed optimization minimizes the errors of each channel independently. Indeed, if there are pixels in a channel with a very high error and pixels in another channel with a very small error, it might result in an average error for the three channels lower than that which would be obtained from a restored image where all pixels of each channel have a similar error. In this context, it is important to note that, for a pixel of the output obtained from a network having only one of the three RGB values correct and the others with a large error, the pixel's color that would be appreciated is different from the expected one. Therefore, the mean square error (\ref{eq:mse}) does not provide a good measure of the quality of the restored image.

To further analyze the existing inaccuracy in the image restoration, we use the supremum norm to measure the difference between each channel of the expected and restored output images. That is, for the expected and restored output images ${{I}^{'}}$ and $Ir$, the error $e_k$ between the $ k $-th channels ${{I_{k}}^{'}}$ and ${Ir}_{k}$, respectively, is computed as
\begin{equation*}\label{eq:supremun_norm}
e_k = \max_{i = 1,\ldots,m,j = 1,\ldots,n } |  I^{'}_{ijk} - {Ir}_{ijk} |. 
\end{equation*} 
The range of these errors for each channel is presented in Table \ref{tab:errores_supremo_R3} for the 22 restored images of the test set. This reveals that, for each image in the test set restored by the network \textbf{R3}, there is at least one pixel whose restoration error is greater than 184, 165, and 189 in the red, green, and blue channels respectively. 

\begin{figure}[h!tb]%
	\begin{center}
		\includegraphics[scale=0.9]{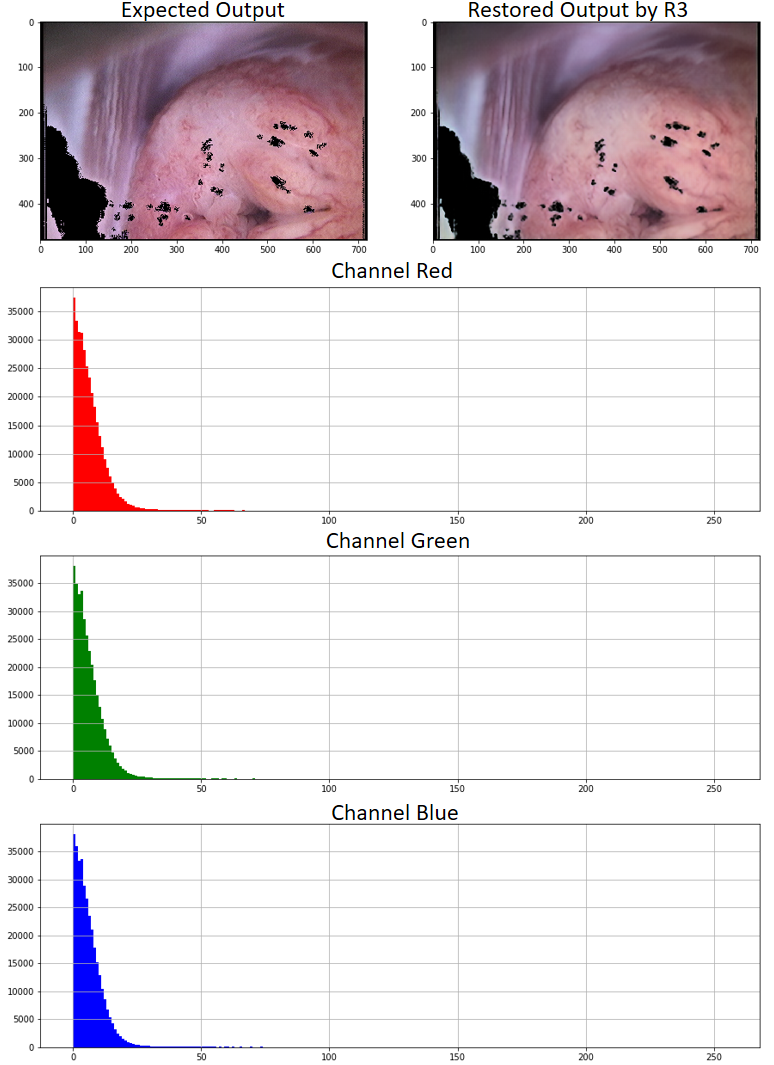}
	\end{center}
	\caption{Histogram of frequency, per channel, of absolute errors between the pixel intensities in the images of the first row. Top left: expected output; Top right:  output obtained by the network \textbf{R3} with the lowest validation error. \label{fig:R3_histograma_de_error_por_canales168-3}}%
\end{figure}

In order to know how frequently these large errors in the restored images appear, the distribution of the absolute errors 
\begin{equation*}\label{eq:absolute_error}
e_{i,j,k} = |  I^{'}_{ijk} - {Ir}_{ijk} |
\end{equation*} of each pixel $(i,j)$ in each channel $k$ of a restored image is analyzed. 
Figure \ref{fig:R3_histograma_de_error_por_canales168-3} plots the histogram of frequency of these errors for each channel of a restored image. It shows that the most frequent error value is 0, and the highest concentration of points is at the beginning of the graph.
\vspace{2cm}

Since the largest absolute error among channels obtained from calculating the supremum norm was 251 for the images of the test set (see Table \ref{tab:errores_supremo_R3}), the range of possible errors among pixels was divided into three intervals, from 0 to 25, from 25 to 50, and from 50 to 251. Table \ref{tab:frec_errores_R3} reports the percentage of pixels with absolute errors in these ranges, for each channel, in the 22 restored images of the test set.
On average, for each channel, at least 95$ \% $ of the pixels in the images restored by \textbf{R3} have an absolute error lower than 25. The green channel tends to be, on average, the channel with the highest percentage of pixels having errors greater than 25. This result demonstrates the good performance of the network \textbf{R3} for restoring the original colors of the test set images.

\subsection{Performance of the selected network to reconstruct unobserved anatomical cervix portion under the SR regions.}\label{sec:Problema_real}

\begin{figure}[h!tb]%
	\begin{center}
		\includegraphics[scale=0.53]{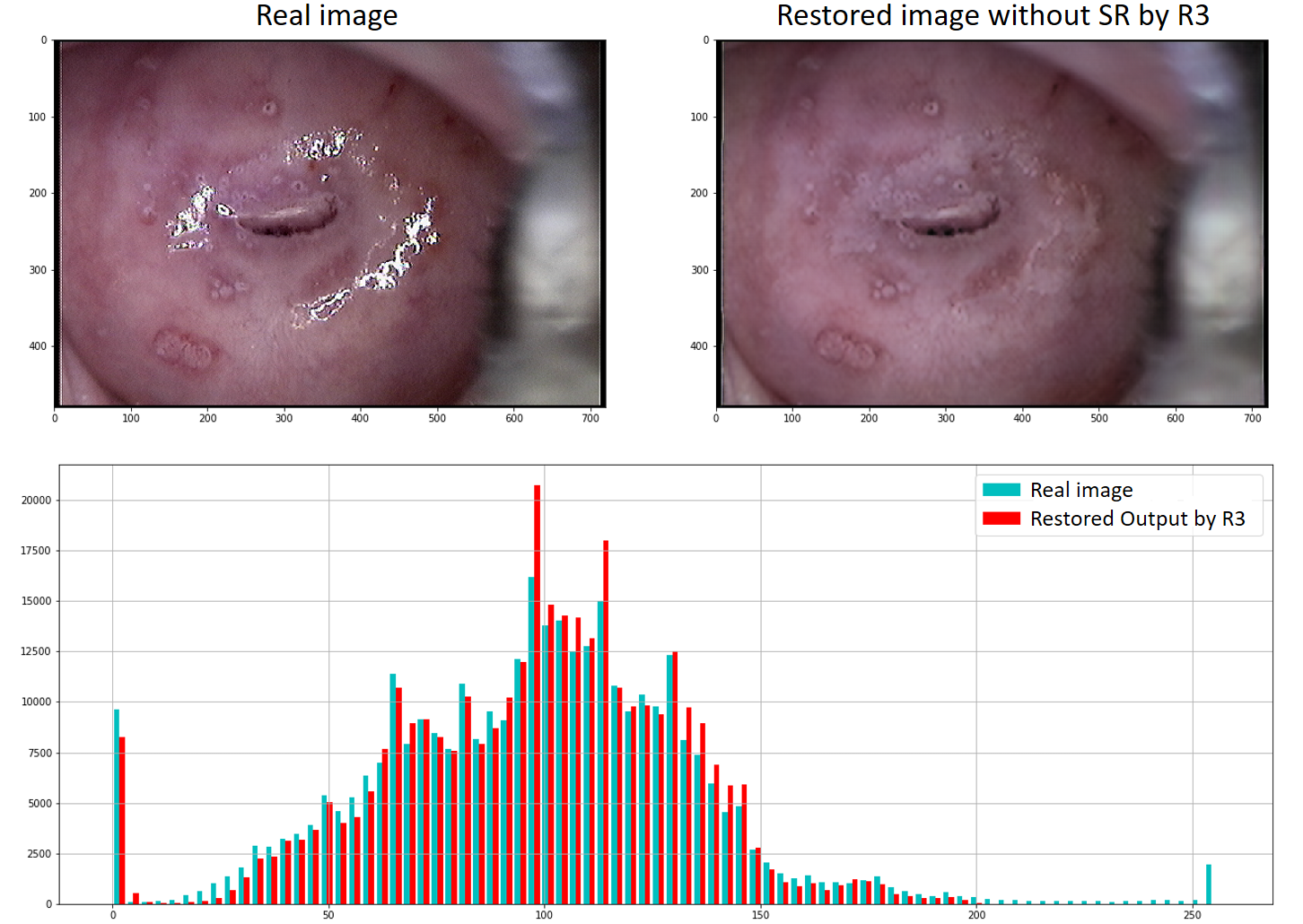}
	\end{center}
	\caption{Result of restoring unobserved anatomical cervix portions under the SR regions using the network with the lowest validation error  \textbf{R3}. Top row, left: real image with SRs; Top row, right: restored image without SRs; Bottom row: histogram of intensities of both images. 		
		\label{fig:histograma1}}%
\end{figure}

Once the trained network \textbf{R3} was selected for having the lowest validation error, and after evaluating its efficiency to restore hidden regions of colposcopic images, it was used to restore the cervix portion under the SR regions. Its performance in this problem was analyzed by a series of experiments on the real test set. 
Figure \ref{fig:histograma1} shows, in the upper part, the image $ I $ and the corresponding restoration $ Ir $ obtained from \textbf{R3}.
The lower part shows the histogram of the color intensities of both images. With color cyan, the real image is represented, and with red color, the network's output.
The histograms illustrate how the pixels with high intensity of the original image disappear in the restored image. It also shows that the distribution of the remaining intensities is reproduced, observing the similarity in both histograms. This behavior was maintained in the 22 analyzed images of the real test set.
\newline

\begin{table}[h!tb]
	\centering
	\scalebox{0.7}{
		\begin{tabular}{|l|l|l|l|}
			\hline\hline
			\backslashbox{Image}{Errors}& Channel Red & Channel Green & Channel  Blue\\ \hline
			\multicolumn{1}{|c|}{Image 1} & \multicolumn{1}{|c|}{208} & 
			\multicolumn{1}{|c|}{186} & \multicolumn{1}{|c|}{210} \\ \hline
			\multicolumn{1}{|c|}{Image 2} & \multicolumn{1}{|c|}{225} & 
			\multicolumn{1}{|c|}{198} & \multicolumn{1}{|c|}{229} \\ \hline
			\multicolumn{1}{|c|}{Image 3} & \multicolumn{1}{|c|}{186} & 
			\multicolumn{1}{|c|}{182} & \multicolumn{1}{|c|}{208} \\ \hline
			\multicolumn{1}{|c|}{Image 4} & \multicolumn{1}{|c|}{224} & 
			\multicolumn{1}{|c|}{193} & \multicolumn{1}{|c|}{229} \\ \hline
			\multicolumn{1}{|c|}{Image 5} & \multicolumn{1}{|c|}{231} & 
			\multicolumn{1}{|c|}{187} & \multicolumn{1}{|c|}{209} \\ \hline
			\multicolumn{1}{|c|}{Image 6} & \multicolumn{1}{|c|}{208} & 
			\multicolumn{1}{|c|}{196} & \multicolumn{1}{|c|}{218} \\ \hline
			\multicolumn{1}{|c|}{Image 7} & \multicolumn{1}{|c|}{194} & 
			\multicolumn{1}{|c|}{237} & \multicolumn{1}{|c|}{208} \\ \hline
			\multicolumn{1}{|c|}{Image 8} & \multicolumn{1}{|c|}{184} & 
			\multicolumn{1}{|c|}{165} & \multicolumn{1}{|c|}{193} \\ \hline
			\multicolumn{1}{|c|}{Image 9} & \multicolumn{1}{|c|}{188} & 
			\multicolumn{1}{|c|}{173} & \multicolumn{1}{|c|}{199} \\ \hline
			\multicolumn{1}{|c|}{Image 10} & \multicolumn{1}{|c|}{202} & 
			\multicolumn{1}{|c|}{213} & \multicolumn{1}{|c|}{213} \\ \hline
			\multicolumn{1}{|c|}{Image 11} & \multicolumn{1}{|c|}{186} & 
			\multicolumn{1}{|c|}{175} & \multicolumn{1}{|c|}{192} \\ \hline
			\multicolumn{1}{|c|}{Image 12} & \multicolumn{1}{|c|}{196} & 
			\multicolumn{1}{|c|}{187} & \multicolumn{1}{|c|}{226} \\ \hline
			\multicolumn{1}{|c|}{Image 13} & \multicolumn{1}{|c|}{198} & 
			\multicolumn{1}{|c|}{182} & \multicolumn{1}{|c|}{235} \\ \hline
			\multicolumn{1}{|c|}{Image 14} & \multicolumn{1}{|c|}{216} & 
			\multicolumn{1}{|c|}{214} & \multicolumn{1}{|c|}{227} \\ \hline
			\multicolumn{1}{|c|}{Image 15} & \multicolumn{1}{|c|}{192} & 
			\multicolumn{1}{|c|}{210} & \multicolumn{1}{|c|}{211} \\ \hline
			\multicolumn{1}{|c|}{Image 16} & \multicolumn{1}{|c|}{238} & 
			\multicolumn{1}{|c|}{228} & \multicolumn{1}{|c|}{251} \\ \hline
			\multicolumn{1}{|c|}{Image 17} & \multicolumn{1}{|c|}{207} & 
			\multicolumn{1}{|c|}{203} & \multicolumn{1}{|c|}{219} \\ \hline
			\multicolumn{1}{|c|}{Image 18} & \multicolumn{1}{|c|}{203} & 
			\multicolumn{1}{|c|}{200} & \multicolumn{1}{|c|}{214} \\ \hline
			\multicolumn{1}{|c|}{Image 19} & \multicolumn{1}{|c|}{199} & 
			\multicolumn{1}{|c|}{192} & \multicolumn{1}{|c|}{196} \\ \hline
			\multicolumn{1}{|c|}{Image 20} & \multicolumn{1}{|c|}{186} & 
			\multicolumn{1}{|c|}{182} & \multicolumn{1}{|c|}{189} \\ \hline
			\multicolumn{1}{|c|}{Image 21} & \multicolumn{1}{|c|}{224} & 
			\multicolumn{1}{|c|}{245} & \multicolumn{1}{|c|}{245} \\ \hline
			\multicolumn{1}{|c|}{Image 22} & \multicolumn{1}{|c|}{190} & 
			\multicolumn{1}{|c|}{198} & \multicolumn{1}{|c|}{206} \\ \hline
			\ \ Average & \ \ \ \ \ 203.8 & \ \ \ \ \ \ 197.0 & \ \ \ \ \ 214.8 \\ 
			\hline
			\ \ Minimum  & \multicolumn{1}{|c|}{184} & \multicolumn{1}{|c|}{165} & 
			\multicolumn{1}{|c|}{189} \\ \hline
			\ \ Maximum & \multicolumn{1}{|c|}{238} & \multicolumn{1}{|c|}{245} & 
			\multicolumn{1}{|c|}{251} \\ \hline\hline
	\end{tabular}}
	\caption{Maximum absolute error in the pixels per channel of each restored image of the test set, using the network \textbf{R3} with the lowest validation error.
	}
	\label{tab:errores_supremo_R3}%
\end{table}

Denote by $ {Int_{max}}^{I} $, $ {Int_{max}}^{'} $ and ${Int_{max}}^{r} $ the maximum intensity corresponding to the original image $ I $, to the modified image $ I^{'}$, and to the restored image $ I_{r} $ obtained by \textbf{R3}. 
${Int_{max}}^{'} $ corresponds to the highest intensity that the algorithm mentioned in Section \ref{sec:Palmer} does not classify as SR in that image. Therefore, if ${Int_{max}}^{'} > {Int_{max}}^{r} $, then all the SRs detected in $ I $ does not appear in $ I_{r} $, and it can be argued that the SRs were removed. 
These three values computed over the images in the real test set are shown in Table \ref{tab:eliminacion_de_brillos}. Clearly, in 21 of the 22 images, the SRs were successfully removed.
It is important to note that this is a criterion for removing SRs, but not for how well the anatomical content under them was estimated.

\begin{figure}[htb]%
	\begin{center}
		\includegraphics[scale=0.5]{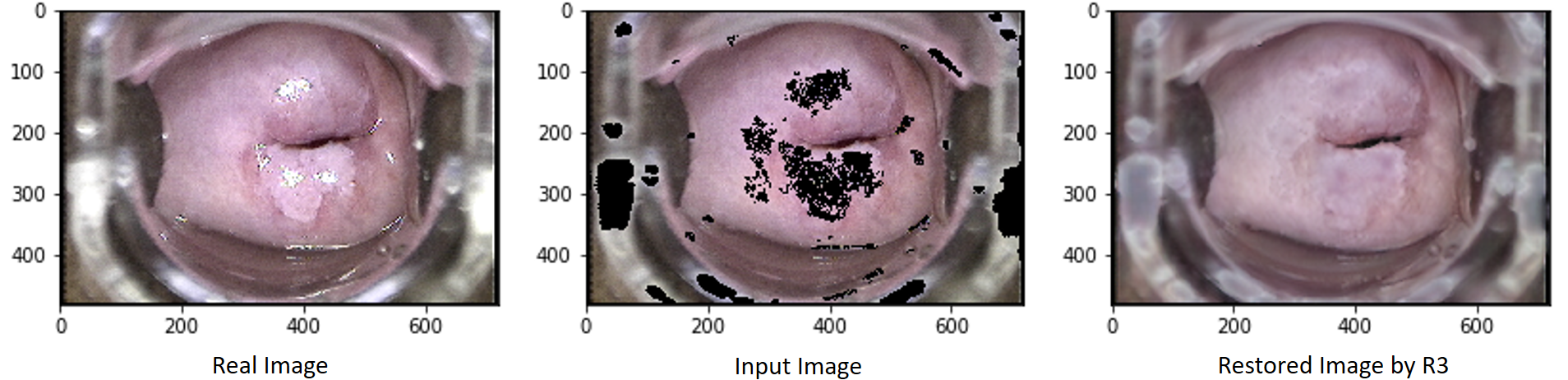}
	\end{center}
	\caption{Effect of the inaccurate SRs detection on the image restoration. From left to right: image of the real test set, input image with detected SR regions bigger than the true ones, and restored image by the network \textbf{R3} with distortions in color and structure. \label{fig:mal_palmer}}%
\end{figure}

\begin{figure}[h!tb]%
	\begin{center}
		\includegraphics[scale=0.5]{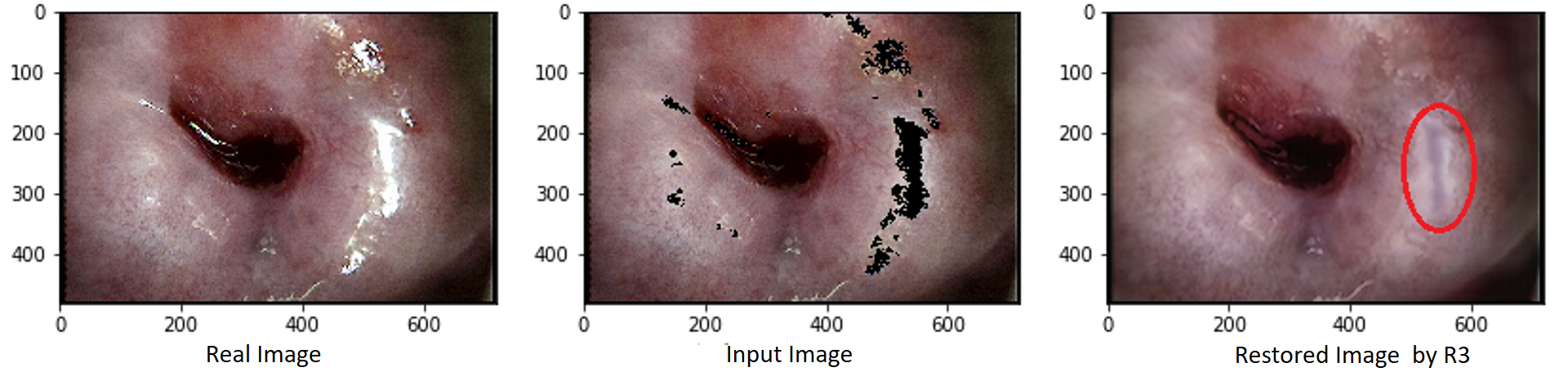}
	\end{center}
	\caption{ Example of undesired  noise appearing in a restored image by \textbf{R3}, indicated in the center of the red circle. \label{fig:ruido_en_restauración}}%
\end{figure}

\begin{figure}[htb]%
	\begin{center}
		\includegraphics[scale=0.5]{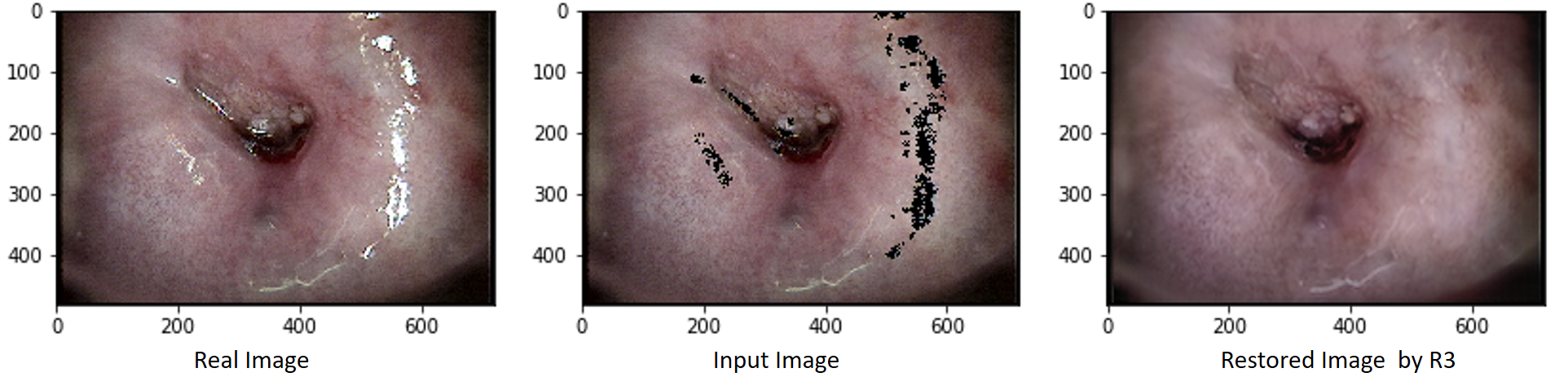}
	\end{center}
	\caption{Example of restored image evaluated by the specialist as satisfactory.  \label{fig:fig_margarita}}%
\end{figure}

For evaluating the restoration of the missing anatomical regions, the expert criterion of a medical specialist was considered. After excluding the three images for which the SRs detection algorithm mentioned in Section \ref{sec:Palmer} does not correctly select the SRs (compromising the quality of the image restoration as shown in Figure \ref{fig:mal_palmer}), the expert analysis of the remainder 19 restored images yields the following conclusion:   $ " $except for the image of Figure  \ref{fig:ruido_en_restauración}, for which appears some noise in the restored area, the brightness in the rest of the reconstructed images was satisfactorily eliminated, allowing the physician to observe the characteristics of the glandular and squamous epithelia of the cervix. In such images, the anatomical elements of the cervix such as glandular orifices (eggs) or Nabothian cyst and physiological characteristics of the cervical mucus are preserved allowing an evaluation of its quality$ "$. An example is shown in Figure \ref{fig:fig_margarita}.

\section*{Conclusions}\label{section:conclusions}
\addcontentsline{toc}{section}{Conclusiones}

In the present work, a neural network-based strategy for specular reflection elimination in colposcopic images was proposed to restore them successfully. A reformulation of the initial problem was done to perform supervised training since the ground truth corresponding to the SR regions is always unknown. The proposed SRs elimination strategy includes the use of an algorithm for SRs identification in colposcopic images, the training of a set of networks to restore any hidden region of colposcopic images, and the use of the network with the lower validation error to restore any unobserved anatomical cervix portion under the SR regions. 
A detailed qualitative and quantitative analysis of the performance of the trained networks shown their capability to restore different hidden regions
of the colposcopic images. The networks with the lowest validation error restore, on average, the 95$\%$ of the pixels in each channel of the images with an error lower than 25 (of a possible maximum of 255). 	

When using the selected network to reconstruct the cervix portion under the SR regions, the brightness was eliminated in 21 of the 22 evaluated images, whereas the distribution of the color intensities of each channel was reproduced, being similar to the expected. 
The restorations of the missing anatomical regions under the SRs were evaluated by a medical expert concluding that -qualitatively- the SRs were satisfactorily eliminated and the gynecological elements of interest were conserved, which facilitates the correct clinical evaluation of the patients.

\begin{table}[htb]
	\centering
	\scalebox{0.74}{	
		\begin{tabular}{|l|l|l|l|l|l|l|l|l|l|}
			\hline\hline
			& \multicolumn{3}{|c|}{Error range 0-25} & \multicolumn{3}{|c|}{Error range 25-50} & \multicolumn{3}{|c|}{Error range 50-251} \\ \cline{2-10}
			& Red & Green & Blue &  Red & Green & Blue &  Red & Green & Blue\\ \hline
			Image 1 & \multicolumn{1}{|c|}{95.6} & \multicolumn{1}{|c|}{96.1} & 
			\multicolumn{1}{|c|}{95.9} & \multicolumn{1}{|c|}{1.8} & 
			\multicolumn{1}{|c|}{1.7} & \multicolumn{1}{|c|}{1.7} & \multicolumn{1}{|c|}{
				2.5} & \multicolumn{1}{|c|}{2.0} & \multicolumn{1}{|c|}{2.3} \\ \hline
			Image 2 & \multicolumn{1}{|c|}{93.6} & \multicolumn{1}{|c|}{94.3} & 
			\multicolumn{1}{|c|}{94.1} & \multicolumn{1}{|c|}{2.6} & 
			\multicolumn{1}{|c|}{2.4} & \multicolumn{1}{|c|}{2.4} & \multicolumn{1}{|c|}{
				3.6} & \multicolumn{1}{|c|}{3.1} & \multicolumn{1}{|c|}{3.4} \\ \hline
			Image 3 & \multicolumn{1}{|c|}{96.3} & \multicolumn{1}{|c|}{96.5} & 
			\multicolumn{1}{|c|}{96.1} & \multicolumn{1}{|c|}{1.7} & 
			\multicolumn{1}{|c|}{1.9} & \multicolumn{1}{|c|}{1.8} & \multicolumn{1}{|c|}{
				1.8} & \multicolumn{1}{|c|}{1.5} & \multicolumn{1}{|c|}{1.9} \\ \hline
			Image 4 & \multicolumn{1}{|c|}{96.5} & \multicolumn{1}{|c|}{97.2} & 
			\multicolumn{1}{|c|}{97.1} & \multicolumn{1}{|c|}{1.6} & 
			\multicolumn{1}{|c|}{1.3} & \multicolumn{1}{|c|}{1.3} & \multicolumn{1}{|c|}{
				1.8} & \multicolumn{1}{|c|}{1.3} & \multicolumn{1}{|c|}{1.5} \\ \hline
			Image 5 & \multicolumn{1}{|c|}{96.8} & \multicolumn{1}{|c|}{97.4} & 
			\multicolumn{1}{|c|}{97.3} & \multicolumn{1}{|c|}{1.4} & 
			\multicolumn{1}{|c|}{1.3} & \multicolumn{1}{|c|}{1.2} & \multicolumn{1}{|c|}{
				1.6} & \multicolumn{1}{|c|}{1.2} & \multicolumn{1}{|c|}{1.4} \\ \hline
			Image 6 & \multicolumn{1}{|c|}{96.4} & \multicolumn{1}{|c|}{96.7} & 
			\multicolumn{1}{|c|}{96.5} & \multicolumn{1}{|c|}{1.8} & 
			\multicolumn{1}{|c|}{1.8} & \multicolumn{1}{|c|}{1.8} & \multicolumn{1}{|c|}{
				1.6} & \multicolumn{1}{|c|}{1.4} & \multicolumn{1}{|c|}{1.5} \\ \hline
			Image 7 & \multicolumn{1}{|c|}{96.5} & \multicolumn{1}{|c|}{93.9} & 
			\multicolumn{1}{|c|}{95.5} & \multicolumn{1}{|c|}{2.3} & 
			\multicolumn{1}{|c|}{4.7} & \multicolumn{1}{|c|}{3.4} & \multicolumn{1}{|c|}{
				1.0} & \multicolumn{1}{|c|}{1.2} & \multicolumn{1}{|c|}{1.0} \\ \hline
			Image 8 & \multicolumn{1}{|c|}{97.8} & \multicolumn{1}{|c|}{98.0} & 
			\multicolumn{1}{|c|}{97.8} & \multicolumn{1}{|c|}{1.1} & 
			\multicolumn{1}{|c|}{1.1} & \multicolumn{1}{|c|}{1.2} & \multicolumn{1}{|c|}{
				1.0} & \multicolumn{1}{|c|}{0.8} & \multicolumn{1}{|c|}{0.9} \\ \hline
			Image 9 & \multicolumn{1}{|c|}{96.6} & \multicolumn{1}{|c|}{97.0} & 
			\multicolumn{1}{|c|}{96.8} & \multicolumn{1}{|c|}{1.7} & 
			\multicolumn{1}{|c|}{1.7} & \multicolumn{1}{|c|}{1.7} & \multicolumn{1}{|c|}{
				1.6} & \multicolumn{1}{|c|}{1.2} & \multicolumn{1}{|c|}{1.4} \\ \hline
			Image 10 & \multicolumn{1}{|c|}{98.4} & \multicolumn{1}{|c|}{98.6} & 
			\multicolumn{1}{|c|}{98.5} & \multicolumn{1}{|c|}{0.8} & 
			\multicolumn{1}{|c|}{0.7} & \multicolumn{1}{|c|}{0.7} & \multicolumn{1}{|c|}{
				0.6} & \multicolumn{1}{|c|}{0.6} & \multicolumn{1}{|c|}{0.7} \\ \hline
			Image 11 & \multicolumn{1}{|c|}{98.3} & \multicolumn{1}{|c|}{98.3} & 
			\multicolumn{1}{|c|}{98.3} & \multicolumn{1}{|c|}{0.6} & 
			\multicolumn{1}{|c|}{0.7} & \multicolumn{1}{|c|}{0.7} & \multicolumn{1}{|c|}{
				0.9} & \multicolumn{1}{|c|}{0.8} & \multicolumn{1}{|c|}{0.9} \\ \hline
			Image 12 & \multicolumn{1}{|c|}{96.9} & \multicolumn{1}{|c|}{96.9} & 
			\multicolumn{1}{|c|}{96.7} & \multicolumn{1}{|c|}{1.3} & 
			\multicolumn{1}{|c|}{1.4} & \multicolumn{1}{|c|}{1.3} & \multicolumn{1}{|c|}{
				1.7} & \multicolumn{1}{|c|}{1.6} & \multicolumn{1}{|c|}{1.8} \\ \hline
			Image 13 & \multicolumn{1}{|c|}{98.3} & \multicolumn{1}{|c|}{98.5} & 
			\multicolumn{1}{|c|}{95.8} & \multicolumn{1}{|c|}{0.9} & 
			\multicolumn{1}{|c|}{0.8} & \multicolumn{1}{|c|}{2.8} & \multicolumn{1}{|c|}{
				0.7} & \multicolumn{1}{|c|}{0.6} & \multicolumn{1}{|c|}{1.3} \\ \hline
			Image 14 & \multicolumn{1}{|c|}{93.6} & \multicolumn{1}{|c|}{94.1} & 
			\multicolumn{1}{|c|}{94.2} & \multicolumn{1}{|c|}{2.8} & 
			\multicolumn{1}{|c|}{3.0} & \multicolumn{1}{|c|}{2.8} & \multicolumn{1}{|c|}{
				3.5} & \multicolumn{1}{|c|}{2.8} & \multicolumn{1}{|c|}{2.9} \\ \hline
			Image 15 & \multicolumn{1}{|c|}{96.1} & \multicolumn{1}{|c|}{95.6} & 
			\multicolumn{1}{|c|}{95.7} & \multicolumn{1}{|c|}{1.9} & 
			\multicolumn{1}{|c|}{2.4} & \multicolumn{1}{|c|}{2.6} & \multicolumn{1}{|c|}{
				1.9} & \multicolumn{1}{|c|}{1.8} & \multicolumn{1}{|c|}{1.6} \\ \hline
			Image 16 & \multicolumn{1}{|c|}{87.9} & \multicolumn{1}{|c|}{85.3} & 
			\multicolumn{1}{|c|}{90.3} & \multicolumn{1}{|c|}{8.0} & 
			\multicolumn{1}{|c|}{10.5} & \multicolumn{1}{|c|}{6.4} & 
			\multicolumn{1}{|c|}{4.0} & \multicolumn{1}{|c|}{4.1} & \multicolumn{1}{|c|}{
				3.2} \\ \hline
			Image 17 & \multicolumn{1}{|c|}{92.8} & \multicolumn{1}{|c|}{90.8} & 
			\multicolumn{1}{|c|}{91.6} & \multicolumn{1}{|c|}{3.1} & 
			\multicolumn{1}{|c|}{5.6} & \multicolumn{1}{|c|}{5.3} & \multicolumn{1}{|c|}{
				4.0} & \multicolumn{1}{|c|}{3.5} & \multicolumn{1}{|c|}{3.0} \\ \hline
			Image 18 & \multicolumn{1}{|c|}{97.1} & \multicolumn{1}{|c|}{97.2} & 
			\multicolumn{1}{|c|}{97.2} & \multicolumn{1}{|c|}{1.2} & 
			\multicolumn{1}{|c|}{1.3} & \multicolumn{1}{|c|}{1.2} & \multicolumn{1}{|c|}{
				1.4} & \multicolumn{1}{|c|}{1.4} & \multicolumn{1}{|c|}{1.5} \\ \hline
			Image 19 & \multicolumn{1}{|c|}{96.9} & \multicolumn{1}{|c|}{97.0} & 
			\multicolumn{1}{|c|}{97.1} & \multicolumn{1}{|c|}{1.6} & 
			\multicolumn{1}{|c|}{1.5} & \multicolumn{1}{|c|}{1.5} & \multicolumn{1}{|c|}{
				1.4} & \multicolumn{1}{|c|}{1.3} & \multicolumn{1}{|c|}{1.2} \\ \hline
			Image 20 & \multicolumn{1}{|c|}{96.7} & \multicolumn{1}{|c|}{96.6} & 
			\multicolumn{1}{|c|}{96.8} & \multicolumn{1}{|c|}{1.5} & 
			\multicolumn{1}{|c|}{1.7} & \multicolumn{1}{|c|}{1.6} & \multicolumn{1}{|c|}{
				1.6} & \multicolumn{1}{|c|}{1.6} & \multicolumn{1}{|c|}{1.5} \\ \hline
			Image 21 & \multicolumn{1}{|c|}{89.6} & \multicolumn{1}{|c|}{85.6} & 
			\multicolumn{1}{|c|}{91.0} & \multicolumn{1}{|c|}{6.6} & 
			\multicolumn{1}{|c|}{10.7} & \multicolumn{1}{|c|}{5.6} & 
			\multicolumn{1}{|c|}{3.6} & \multicolumn{1}{|c|}{3.6} & \multicolumn{1}{|c|}{
				3.2} \\ \hline
			Image 22 & \multicolumn{1}{|c|}{97.2} & \multicolumn{1}{|c|}{97.2} & 
			\multicolumn{1}{|c|}{97.0} & \multicolumn{1}{|c|}{1.4} & 
			\multicolumn{1}{|c|}{1.5} & \multicolumn{1}{|c|}{1.4} & \multicolumn{1}{|c|}{
				1.2} & \multicolumn{1}{|c|}{1.2} & \multicolumn{1}{|c|}{1.5} \\ \hline\hline
			\ \ Average & \multicolumn{1}{|c|}{95.7} & \multicolumn{1}{|c|}{95.4} & 
			\multicolumn{1}{|c|}{95.8} & \multicolumn{1}{|c|}{2.2} & 
			\multicolumn{1}{|c|}{2.7} & \multicolumn{1}{|c|}{2.3} & \multicolumn{1}{|c|}{
				2.0} & \multicolumn{1}{|c|}{1.8} & \multicolumn{1}{|c|}{1.8} \\ \hline
			\ \ Minimum & \multicolumn{1}{|c|}{87.9} & \multicolumn{1}{|c|}{85.3} & 
			\multicolumn{1}{|c|}{90.3} & \multicolumn{1}{|c|}{0.6} & 
			\multicolumn{1}{|c|}{0.7} & \multicolumn{1}{|c|}{0.7} & \multicolumn{1}{|c|}{
				0.6} & \multicolumn{1}{|c|}{0.6} & \multicolumn{1}{|c|}{0.7} \\ \hline
			\ \ Maximum & \multicolumn{1}{|c|}{98.4} & \multicolumn{1}{|c|}{98.6} & 
			\multicolumn{1}{|c|}{98.5} & \multicolumn{1}{|c|}{8.0} & 
			\multicolumn{1}{|c|}{10.7} & \multicolumn{1}{|c|}{6.4} & 
			\multicolumn{1}{|c|}{4.0} & \multicolumn{1}{|c|}{4.1} & \multicolumn{1}{|c|}{
				3.4} \\ \hline
			\ \ Median & \multicolumn{1}{|c|}{96.6} & \multicolumn{1}{|c|}{96.8} & 
			\multicolumn{1}{|c|}{96.6} & \multicolumn{1}{|c|}{1.6} & 
			\multicolumn{1}{|c|}{1.7} & \multicolumn{1}{|c|}{1.7} & \multicolumn{1}{|c|}{
				1.6} & \multicolumn{1}{|c|}{1.4} & \multicolumn{1}{|c|}{1.5} \\ \hline\hline
		\end{tabular}
	}
	\caption{Percentage of pixels per channel within the specified absolute error ranges, for each image of the test set restored by the network \textbf{R3}. }
	\label{tab:frec_errores_R3}%
\end{table}

\begin{table}[htb]
	\centering
	\scalebox{0.84}{
	\begin{tabular}{|c|c|c|c|c|}
		\hline
		& $ {Int_{max}}^{I} $ & $ {Int_{max}}^{'} $ &$ {Int_{max}}^{r} $ & ${Int_{max}}^{'} > {Int_{max}}^{r} $\\ \hline
		Image	1  &  255.0  &  216.7  &  200.0 & Yes \\ \hline
		Image	2  &  255.0  &  216.7  &  206.3 & Yes\\ \hline
		Image	3  &  255.0  &  216.7  &  189.0 & Yes\\ \hline
		Image	4  &  255.0  &  216.3  &  205.3 & Yes\\ \hline
		Image	5  &  255.0  &  216.0  &  208.7 & Yes\\ \hline
		Image	6  &  255.0  &  216.0  &  198.3 & Yes\\ \hline
		Image	7  &  255.0  &  216.0  &  207.0 & Yes\\ \hline
		Image	8  &  255.0  &  215.3  &  199.7 & Yes\\ \hline
		Image	9  &  255.0  &  216.0  &  196.0 & Yes\\ \hline
		Image	10  &  248.0  &  210.3  &  197.3 & Yes \\ \hline
		Image	11  &  253.0  &  213.3  &  186.0 & Yes\\ \hline
		Image	12  &  255.0  &  216.7  &  202.7 & Yes\\ \hline
		Image	13  &  255.0  &  216.7  &  197.7 & Yes\\ \hline
		Image	14  &  255.0  &  216.3  &  204.0 & Yes\\ \hline
		Image	15  &  255.0  &  216.3  &  192.7 & Yes\\ \hline
		Image	16  &  255.0  &  216.7  &  208.3 & Yes\\ \hline
		Image	17  &  254.0  &  215.7  &  198.3 & Yes\\ \hline
		Image	18  &  255.0  &  216.3  &  204.0 & Yes\\ \hline
		Image	19  &  255.0  &  216.7  &  197.0 & Yes\\ \hline
		Image	20  &  255.0  &  216.7  &  179.3 & Yes\\ \hline
		Image	21  &  255.0  &  216.7  &  209.7 & Yes\\ \hline
		Image	22  &  239.7  &  203.7  &  212.0 & No \\ \hline
	\end{tabular}
}
\caption{ Maximum intensity values of the original image $ I $, of the image $ I^{'}$ resulting from blacking out the pixels selected by the real mask, and of the image $ I_{r}$ obtained by the network \textbf{R3}. Last column indicates when the SRs were removed from the images.}
	\label{tab:eliminacion_de_brillos}%
\end{table}

\vspace{2cm}

\bibliographystyle{plain}
\bibliography{Specular_Reflections_Removal}

\end{document}